%% file: europar.tex
\documentclass[11pt,twoside,a4paper]{article}
\usepackage{makeidx}
\usepackage{authblk}
\usepackage{amsfonts}
\usepackage{amsmath}
\usepackage[dvips]{graphicx}
\usepackage{url}
\usepackage{multirow}
\usepackage{setspace}
\usepackage{verbatim}
\usepackage{algorithm}
\usepackage{algorithmicx}
\usepackage{algpseudocode}
\usepackage{subfig}

\begin{document}

\title{First Experiences Optimizing Smith-Waterman on Intel's Knights Landing Processor}
\author[1]{Enzo Rucci\thanks{erucci@lidi.info.unlp.edu.ar}}
\author[1]{Carlos Garcia\thanks{garsanca@ucm.es}}
\author[1]{Guillermo Botella\thanks{gbotella@ucm.es}}
\author[2]{Armando De Giusti}
\author[3]{Marcelo Naiouf}
\author[1]{Manuel Prieto-Matias}
\affil[1]{III-LIDI, CONICET, Facultad de Inform\'atica, Universidad Nacional de La Plata}
\affil[2]{Depto. Arquitectura de Computadores y Autom\'atica, Universidad Complutense de Madrid}
\affil[3]{III-LIDI, Facultad de Inform\'atica, Universidad Nacional de La Plata}

\maketitle

%-------------------------------------------------------------------------------

\begin{abstract}
The well-known Smith-Waterman (SW) algorithm is the most commonly used method for local sequence alignments. However, SW is very computationally demanding for large protein databases. There exist several implementations that take advantage of computing parallelization on many-cores, FPGAs or GPUs, in order to increase the alignment throughtput. In this paper, we have explored SW acceleration on Intel KNL processor. The novelty of this architecture requires the revision of previous programming and optimization techniques on many-core architectures. To the best of authors knowledge, this is the first KNL architecture assessment for SW algorithm. Our evaluation, using the renowned Environmental NR database as benchmark, has shown that multi-threading and SIMD exploitation reports competitive performance (351 GCUPS) in comparison with other implementations.

\end{abstract}

\input{intro}

\input{sw}
\input{phi}

\input{implementation}

\input{results}

\input{conclusions}

% use section* for acknowledgement
\subsubsection*{Acknowledgments}
This work has been partially supported by Spanish government through research contract TIN2015-65277-R and CAPAP-H5 network (TIN2014-53522).

%-------------------------------------------------------------------------------
\bibliographystyle{plain}
\bibliography{biblio}

\end{document}

%% file: intro.tex
\section{Introduction}
\label{sec:intro}

Nowadays the greatest challenge of Bioinformatics is no longer data generation but also efficient information analysis and interpretation. In fact, sequencing technologies~\cite{Mount04} is currently considered one of the most successful instruments in Bioinformatics, basically solved by heuristic methods.

The key aspect of Smith-Waterman (SW) algorithm~\cite{Smith1981} is that always finds the optimal local alignment between two sequences. This characteristic makes this method the basis of more sophisticated alignment technologies, so its study and acceleration in different platforms has motivated a great interest for the scientific community. Although, many approaches, such as BLAST~\cite{blast} and FASTA~\cite{fasta} are more efficient in term of execution time, they do not guarantee the optimal alignment. 

SW establishes similar regions between two DNA or protein sequences. A score matrix must be built in order to determine the best alignment. Besides, matrix size depends on sequence lengths which determines the parallel scalability. From parallel processing perspective, regarding DNA alignment with sequences up to hundreds of million nucleotide, the huge matrix created only permits to perform a single sequence pair, so the low-level parallelism available in the alignment can be exploited by means of the \emph{intra-task} scheme. Nevertheless, protein sequences which are shorter requires small matrices. This aspect permits to exploit coarse level parallelism computing multiple independent alignments simultaneously in \emph{inter-task} approach way.

%Related work
The computational complexity of the SW algorithm has motivated a large amount of research in order to reduce execution time by means of acceleration on a great variety of architectures. In the last years in the context of SW protein alignment, we have witnessed SIMD-vector exploitation~\cite{SW-SIMD-Farrar,SW-Rognes-2000,Rognes2011} available now on modern CPUs, highlighting the recently released \emph{Parasail} library~\cite{Parasail}. In the field of heterogeneous computing, the most successful solution is the \textit{CUDASW++} software~\cite{cudasw-3} 
%, which achieve up to 185.6 GCUPS (billion cell updates per second) 
for multi CUDA-enabled Graphics Processor Units (GPUs) with concurrent CPU computing. Moreover, for Intel's co-processors based on Xeon Phi, we highlight both optimized hand-tuned SW implementations denominated as \emph{SWAPHI}~\cite{SWAPHI} and \emph{LSBDS}~\cite{LSBDS}.  
%which obtain up to 228 GCUPS and 220 GCUPS respectively with different input databases. 
Besides centering on Intel Xeon Phi alternative, Rucci et al.~\cite{SW-CPE15} have recently studied also energy efficiency on a hybrid implementation that exploits both CPU and co-processors simultaneously. Using FPGAs as accelerators, we can found linear systolic array implementations for Xilinx Virtex FPGAs~\cite{SW-FPGA-Isa,SW-FPGA-Oliver}, custom instructions~\cite{SW-FPGA-Li} and the proposal of Rucci \emph{et al.}~\cite{OSWALD16} where the behavior of the novel paradigm of OpenCL on Altera's FPGAs is studied, whose most relevant results show that these devices are the most efficient from energy footprint perspective.

Our paper proposes and evaluates a SW algorithm using the last generation of Intel's Xeon Phi with the Knights Landing (KNL) architecture. We would like to note that although there exist SW studies in old Xeon Phi with Knights Corner (KNC) architecture~\cite{SWAPHI,LSBDS,SW-CPE15}, to the best of authors knowledge there are no related works in Bioinformatics scenario with KNL architecture due to its recent commercialization. Among the main differences of KNL respect to its predecessor, are the incorporation of AVX-512 extensions, a remarkable number of vector units increment and new on-package high-bandwidth memory. These aspects make necessary the revision of the previous optimization proposals for the SW algorithm.

% Contenido del paper The rest of the paper is organized as follows. 
Section~\ref{sec:SW-Algorithm} introduces the basic concepts of the Smith-Waterman algorithm. Section~\ref{sec:xeon_phi} briefly introduces the Intel's Xeon Phi architecture and in Section~\ref{sec:implementation} we describe our implementation of the SW algorithm. In Section~\ref{sec:results} we discuss performance results and finally in Section~\ref{sec:conclusion} we conclude with some ideas for future research.

%% file: sw.tex
\section{Smith-Waterman Algorithm}
\label{sec:SW-Algorithm}

%The SW algorithm is used to obtain the optimal local alignment between two sequences and was proposed by Smith and Waterman~\cite{Smith1981}. This method employs a dynamic programming approach and its high sensitivity comes from exploring all the possible alignments between two sequences. 

Given two sequences $S_1$ and $S_2$, with sizes $|S_1|=m$ and $|S_2|=n$, the recurrence relations for the SW algorithm with affine gap penalties~\cite{gotoh81} are defined below.

\begin{equation}
	H_{i,j}=max \{0, H_{i-1,j-1}+SM(S_1[i],S_2[j]), E_{i,j}, F_{i,j}\}
	\label{eq:sw2}
\end{equation}

\begin{equation}
	E_{i,j}=max \{H_{i,j-1} - G_{oe}, E_{i,j-1} - G_{e}\}
	\label{eq:sw3}
\end{equation}
	
\begin{equation}
	F_{i,j}=max \{H_{i-1,j} - G_{oe}, F_{i-1,j} - G_{e}\}
	\label{eq:sw4}
\end{equation}	

$H_{i,j}$ contains the score for aligning the prefixes $S_1[1..i]$ and $S_2[1..j]$. $E_{i,j}$ and $F_{i,j}$ are the scores of prefix $S_1[1..i]$ aligned to a gap and prefix $S_2[1..j]$ aligned to a gap, respectively. \emph{SM} is the \emph{scoring matrix} which defines the substitution scores for all residue pairs. Generally $SM$ rewards with a positive value when $q_{i}$ and $d_{j}$ are identical or relatives, and punishes with a negative value otherwise. $G_{oe}$ is the sum of gap open and gap extension penalties while $G_{e}$ is the gap extension penalty. The recurrences should be calculated with $1 \leq i \leq m$ and $1 \leq j \leq n$, after initializing $H$, $E$ and $F$ with 0 when $i = 0$ or $j = 0$. The maximum value in the alignment matrix $H$ is the optimal local alignment score.

\begin{figure}
	\begin{centering}
	\includegraphics[width=0.4\columnwidth]{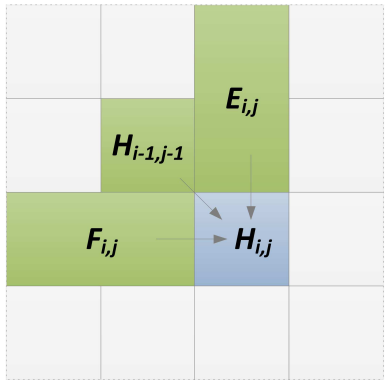}
	\par
	\end{centering}
	\caption{\label{fig:dependences}Data dependences to compute \emph{H}.}
\end{figure}

It is important to note that \emph{H} values can not be computed in any order due to the data dependences inherent to this problem. To be able to calculate the value of any cell, all the values of the previous
 cells at the same row and column have to be computed first, as shown in Figure~\ref{fig:dependences}.
 These dependences restrict the ways in that \emph{H} can be computed.

%% file: phi.tex
\section{Intel's Xeon Phi}
\label{sec:xeon_phi}

With the Exascale challenge as a target in High Performance Computing (HPC), accelerators seem to be the alternative to achieve such goals due to consumption constrains in general-purpose processors. Xeon Phi (Phi) is the code brand name given by Intel to a series of massively many-core processor designed for HPC purposes. Phi corresponds to an specialized architecture denominated as Intel Many Core Architecture (MIC) in contract to Multi-Core Architecture for General-purpose processors. Phi architecture derived from the defunct Larrabee project~\cite{Larrabee} and the Teraflops Research Chip research project. In 2012, Intel launches the first Phi generation (KNC) which main features up to 61 x86 pentium cores with extended vector units (512-bit) and simultaneous multithreading (four hardware threads per core). Meanwhile first Phi was attached to the host processor via PCI Express bus, second generation (KNL) can operate as standalone processor.

\begin{figure}[tb]
	\begin{centering}
	\includegraphics[width=0.9\columnwidth]{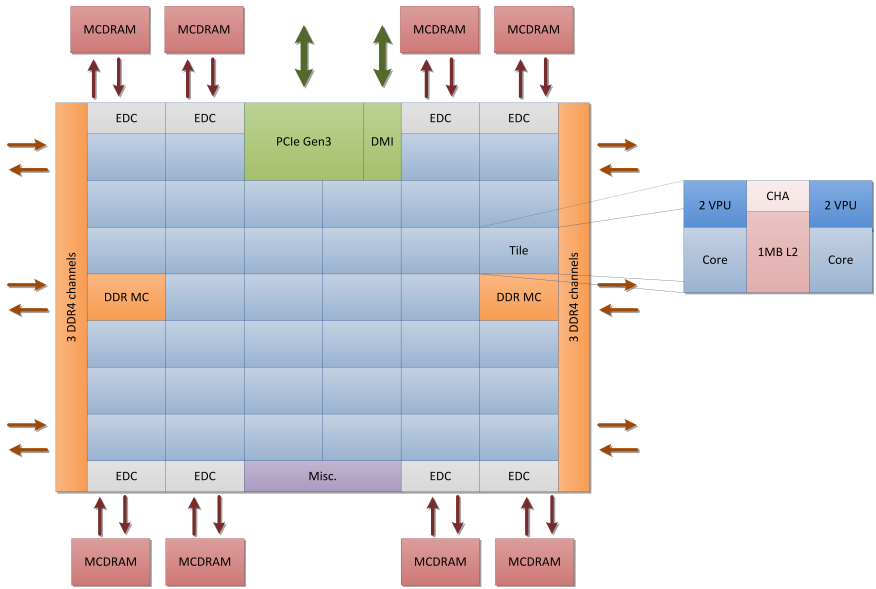}
	\par
	\end{centering}
	\caption{\label{fig:phifig} Xeon Phi KNL architecture.}
\end{figure}

As shown in Figure~\ref{fig:phifig}, KNL architecture corresponds up to 36 \emph{Tiles} interconnected by 2D mesh. Each Tile includes 2 cores based on the out-of-order Intel's Atom micro-architecture (4 threads per core), 2 Vector Processing Units (VPUs) with AVX-512 support and a shared L2 cache of 1 MByte. 

One of the main differences of the KNL architecture with respect to its predecessor is the availability of on-package high-bandwidth memory (HBM). This particular technology permits three configuration modes: \emph{Flat mode}, \emph{Cache Mode} and \emph{Hybrid mode}. While in \emph{Cache mode}, HBM is used as classical cache with lower performance rates and null source code changes, in \emph{Flat mode} the HBM is used as addressable memory being necessary the programmer intervention to indicate manually which part of its source code is allocated to HBM. It is important to note that in \emph{Flat mode}, MCDRAM is treated as Non-Uniform-Memory-Access architectures (NUMA), thus programmer should take special care for achieving efficient memory access from the cores~\cite{IntelMCDRAM}.

KNL supports not only old Intel's multimedia extensions such as 128-bit SSE\emph{x} and 256-bit AVX\emph{x}, but also modern 512-bit AVX-512. In fact, Intel will unify the SIMD instruction-set on both general purpose (announce its support on Xeon E5-26\emph{xx} V5 at 2017) and KNL processors by means of AVX-512. AVX-512 performs 512-bit SIMD capabilities, 32 logical registers, vector predication via eight new mask registers and gather/scatter indirect vector accesses. Currently, modern Phi has two VPUs per core allowing SIMD parallelism which acts as 32 SIMD-lanes for single-precision (512 bits registers/32 bits in SP $\times$ 2 VPUs = 32 lanes) and 16 SIMD-lanes for double-precision~\cite{KNL_micro2016}. Although Intel AVX-512 instructions contains several categories, Xeon Phi KNL architecture only supports four: AVX-512F (foundation instructions); AVX-512CD (conflict-detection); AVX-512ER (exponential and reciprocal); and  AVX-512PF (prefetch instructions). 

From a programming point of view, one of the main goals of this platform is the support of existing parallel programming models traditionally used on HPC scenario such as the OpenMP, MPI or TBB paradigms~\cite{KNLbook}, which simplifies code development and improves portability over other alternatives based on accelerator-specific programming languages such as CUDA or OpenCL. In fact, although it should not be the most efficient way, KNL allows binary compatibility with Xeon families.

However, minimal programming efforts such as the introduction of some directives to inform the compiler about pointer disambiguation or data alignment data dependencies usually provide poor performance rates. In fact, \textit{guided auto-vectorization} is not able to achieve the best performance in most cases and programmers usually need to make an extra effort to hand-tune the codes to exploit SIMD capabilities. Indeed, intrinsics are currently the only option for complex applications which suffer from data dependencies or irregular access patterns that can be hidden using specific code transformations. Unfortunately, improving performance comes at the expense of losing cross-platform portability. %Most processor families, even from the same vendor, have non-compatible intrinsics which support different SIMD instruction sets. As a consequence, code developers need to write many code branches, thus increasing maintenance needs.

%% file: implementation.tex
\section{SW Implementation}
\label{sec:implementation}

In this section we will address the optimizations performed on the Intel Xeon Phi KNL processor. Before describing them in detail, we would like to point out the algorithm flow which can be summarized in the following steps:
\begin{enumerate}
\item \textit{Pre-processing stage}: database sequences are pre-processed to allow subsequent parallel computation.
\item \textit{SW stage}: alignments are carried out.
\item \textit{Sorting stage}: alignment scores are sorted in descending order.
\end{enumerate}

The inter-task parallelism approach is performed in order to exploit the SIMD vector capabilities available on the Xeon Phi KNL processor. In that sense, database sequences are processed in groups and the size of the groups is determined by the number of SIMD vector lanes.  Before grouping sequences, database sequences are sorted by their lengths in ascending order and padded with dummy symbols. This is done to favor memory pattern access and minimize workload imbalances. 
%Because this process must be repeated for every search, sequence databases are preprocessed separately to avoid duplicate work. The databases are read from the FASTA format\footnote{FASTA format description: \url{http://blast.ncbi.nlm.nih.gov/blastcgihelp.shtml}} and then transformed to a binary format. In addition, this conversion allows a subsequent faster reading of sequences to main memory.

\subsection{Multiple Parallelism Levels}

\begin{figure}[tb]
    \centering
    \begin{minipage}{0.48\textwidth}
        \centering
        \includegraphics[width=1\textwidth]{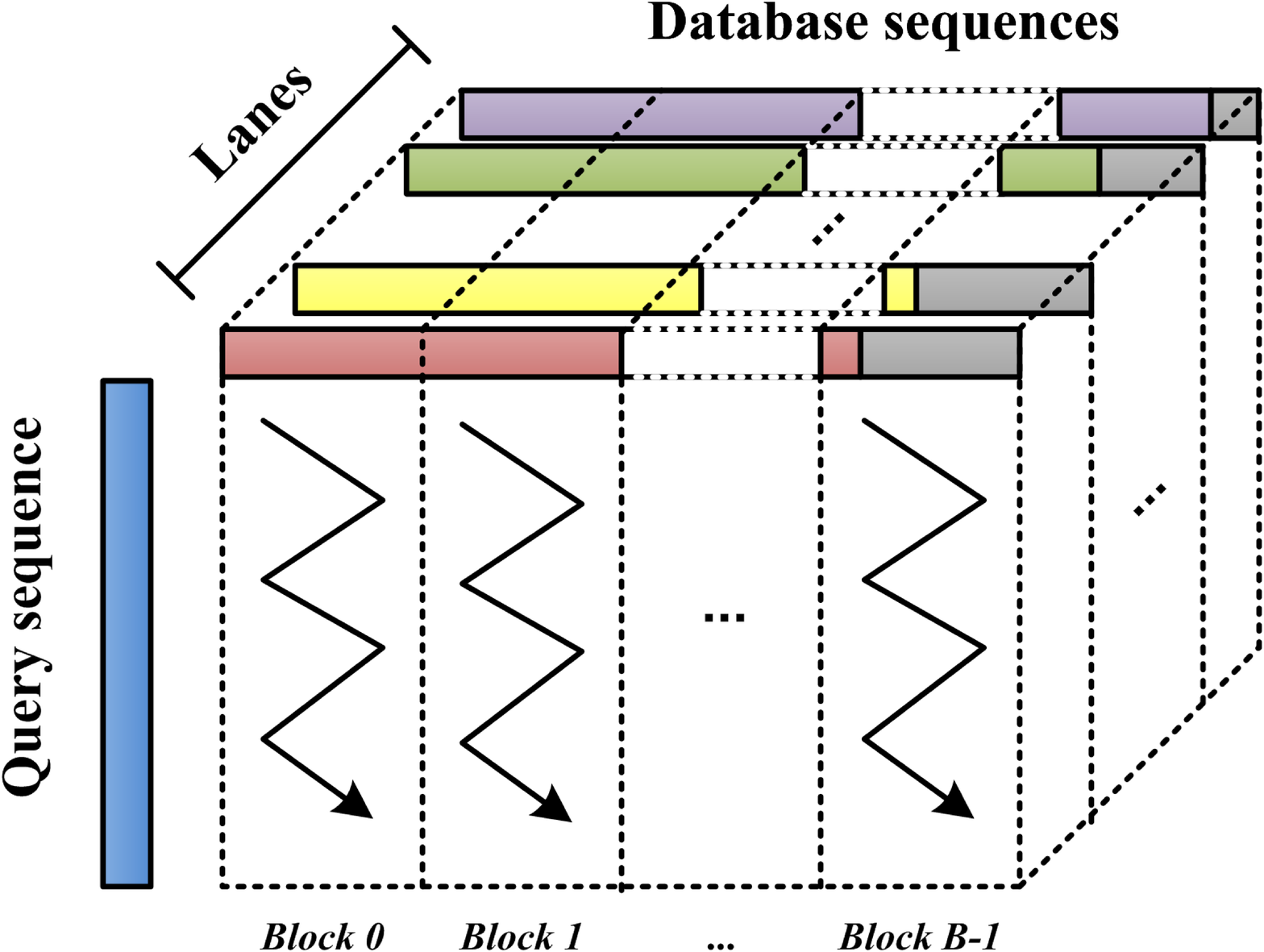} % first figure itself
        \caption{\label{fig:kernel}Schematic representation of the inter-task matrix computation}
    \end{minipage}\hfill
    \begin{minipage}{0.48\textwidth}
        \centering
        \includegraphics[width=1\textwidth]{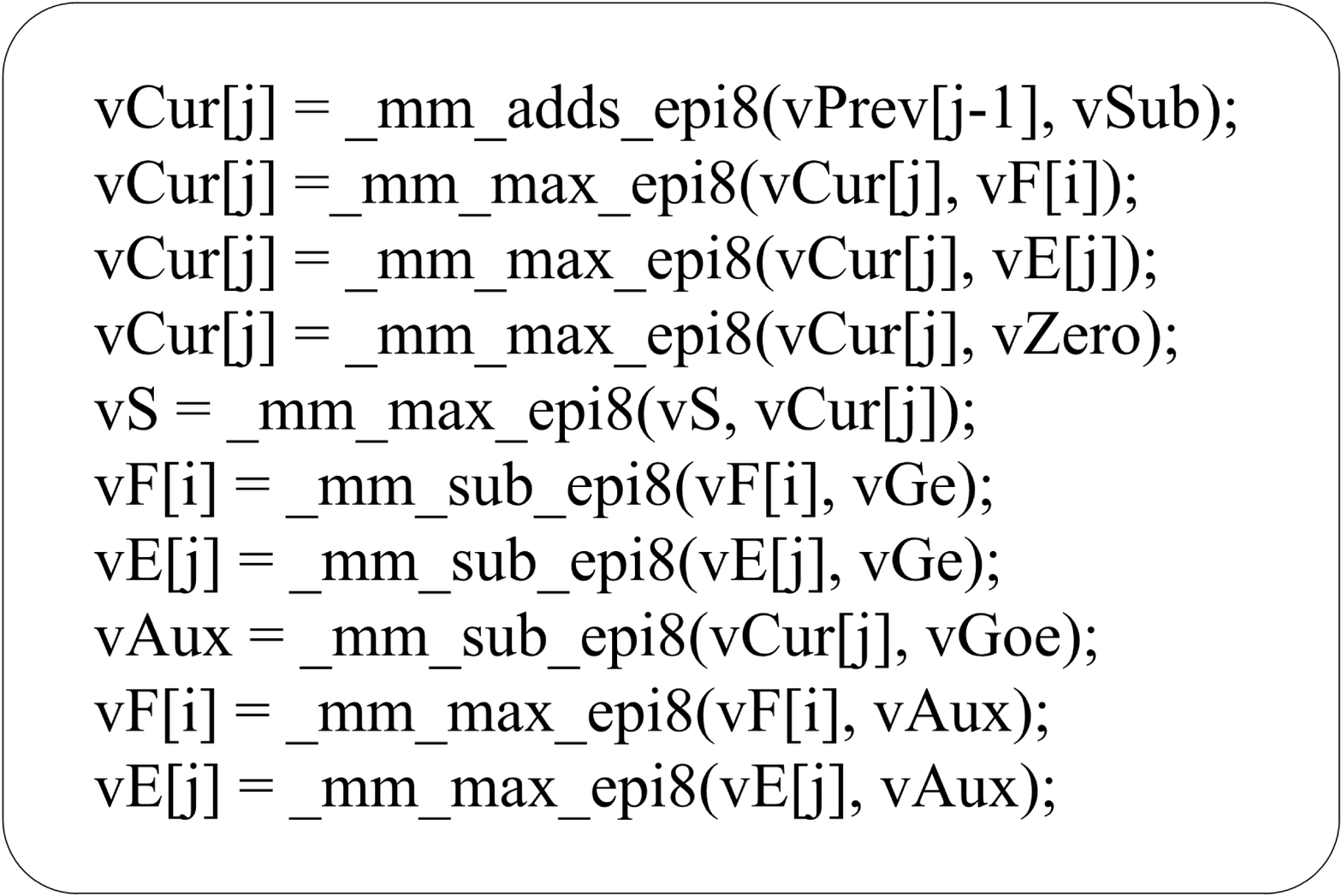} % second figure itself
        \caption{\label{fig:sse} SSE4.1 core instructions}
    \end{minipage}
\end{figure}

Our implementation exploits both data and thread parallelism levels. On one hand, we have used SIMD instructions by means of hand-tuned intrinsic functions. In particular, we have explored the usage of SSE4.1, AVX2 and AVX-512 extensions. On the other hand, we take advantage of the OpenMP programming model to express parallelism across multiple cores.
% decía "thorough", quisieron decir "through"?
The database sequences are dinamically distributed among the cores as soon as the threads become idle. Each alignment matrix is divided into vertical blocks and computed in a row-by-row manner (see Figure~\ref{fig:kernel}). This blocking technique improves data locality reducing the number of cache misses. In addition, the inner loop is fully unrolled to increase performance.

\begin{figure}[tb]
    \centering
    \begin{minipage}{0.48\textwidth}
        \centering
        \includegraphics[width=1\textwidth]{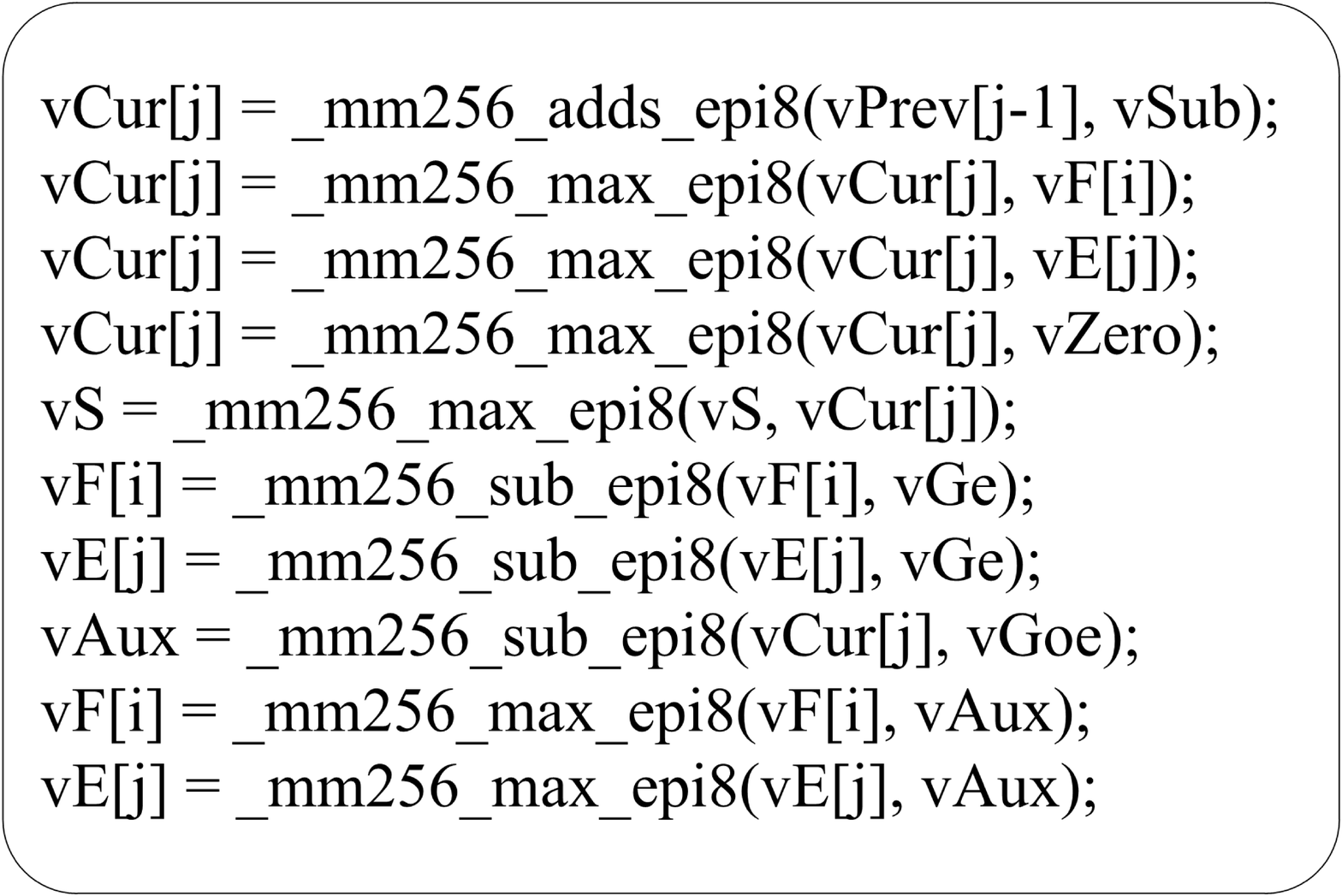} % second figure itself
        \caption{\label{fig:avx2} AVX2 core instructions}
    \end{minipage}\hfill
    \begin{minipage}{0.48\textwidth}
        \centering
        \includegraphics[width=1\textwidth]{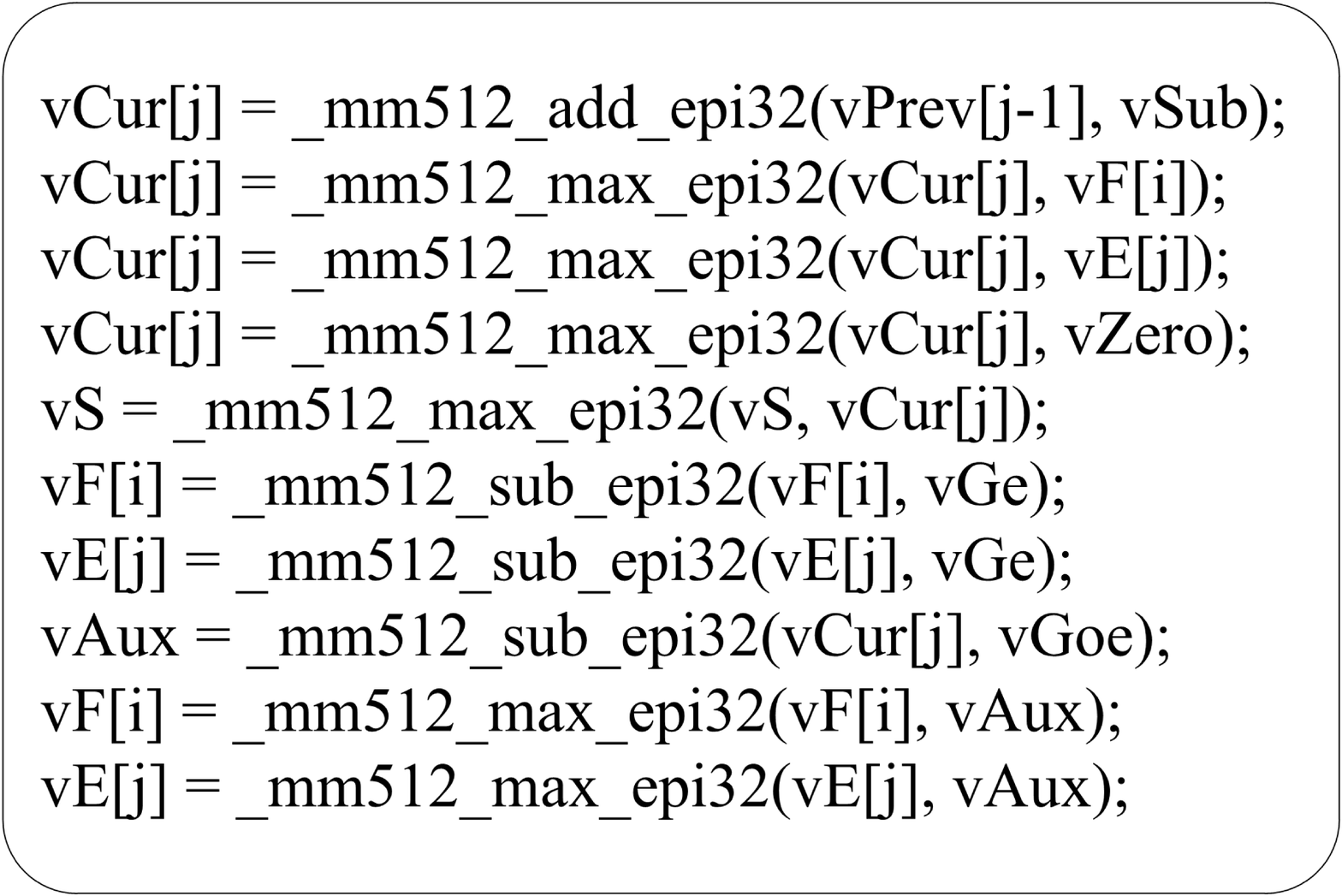} % second figure itself
        \caption{\label{fig:avx-512} AVX-512 core instructions}
    \end{minipage}
\end{figure}

Figures~\ref{fig:sse}, \ref{fig:avx2} and ~\ref{fig:avx-512} show the core instructions of SSE4.1, AVX2 and AVX-512 extensions, respectively. $vCur$ is the block row being calculated while $vPrev$ is the previous one. After computing the current block row, $vCur$ and $vPrev$ are swapped to process the next row. Besides, $vSub$ represents the substitution scores for the database sequence residues against the query residue. $vE$ and $vF$ are the score vectors for alignments ending in a gap in the query and the database sequence, respectively. $vGoe$ represents the vector for the sum of gap open and gap extension penalties while $vGe$ is the vector for gap extension penalty. Last, $vS$ keeps the current optimal alignment score. 
%\input{alg1}

%acá poner algo de que se intercambian los buffers

\subsection{Instruction Set and Integer Range Selection}
\label{sec:integer-range}

Although almost all alignment scores can be represented using an 8-bit integer range in order to express as much SIMD parallelism available, there are some alignments that can not be expressed with this integer range so a wider range should be used. In the context of KNL processors, it is supported SSE\emph{x}, AVX\emph{x} and AVX-512 instructions sets. While SSE4.1 extensions allow computation of 16 alignments in parallel, AVX2 instructions double this number. Saturated arithmetic operations are used in additions operation to detect overflow computation. When potential overflow is detected (i.e. the alignment score is equal to the maximum value of the integer representation employed), the alignment is recalculated using the next wider integer range. Overflow checking is performed to verify if overflow occurred in the lower/upper half or in both halves of the score vector in order to avoid unnecessary recalculations. Unfortunately Xeon Phi KNL processors do not include AVX-512BW subset (byte and word version of instructions in AVX-512F). This fact means that the narrowest integer range in these devices is 32 bit for AVX-512. So AVX-512 cannot compute more alignments simultaneously than SSE4.1 or AVX2. In contrast, operations for overflow detection are not required.

\subsection{Substitution scores}
Our code also implements other well-known optimizations of the SW algorithm that have been proposed in previous works, such as the Query Profile (\emph{QP})~\cite{SW-Rognes-2000} and Score Profile (\emph{SP})~\cite{Rognes2011} optimisations. 
\begin{itemize}
	\item The \emph{QP} strategy is based on creating an auxiliary two-dimensional array of size $|q|\times|\sum|$, where $q$ is the query sequence and $\sum$ is the alphabet. Each row of this array contains the scores of the corresponding query residue against each possible residue in the alphabet. Since each thread compares the same query residue against different ones from the database, this optimization improves data locality at the cost of a negligible increment in memory requirements.
	\item The \emph{SP} technique is based on constructing an auxiliary $n \times L \times \sum$  score array, where $n$ is the length of the database sequence, $L$ is the number of vector lanes and $\sum$ is the alphabet. This auxiliary structure contains the substitutions scores for each query-database residue combination and is constructed before matrix computation. Since each row of the \emph{SP} forms an $L$-lane score vector, an advantage is that its values can be gathered using a single vector load reducing the number of operations in the innermost loop. However, because the \emph{SP} must be re-built for each database sequence, its suitability must be evaluated, especially for short queries.
\end{itemize}

%% file: results.tex
\section{Experimental Results}
\label{sec:results}

\subsection{Experimental Design}
All tests have been performed on an Intel server running CentOS 7.2 equipped with a Xeon Phi 7250 processor 68-core 1.40GHz (4 hw thread per core and 16GB HBW memory) and 64GB main memory. The processor was run in \emph{Flat} memory mode and \emph{Quadrant} cluster mode.

We have used Intel's ICC compiler (version 17.0.1.132) with the \emph{-O3} optimization level by default. 
The experiments used to assess performance are similar to those in previous work~\cite{OSWALD16,SW-CPE15,cudasw-3,Rognes2011}. We have evaluated our application by searching 20 query protein sequences against the well-known  Environmental NR database (release 2016\_11)~\footnote{The Environmental NR database is available online at \url{ftp://ftp.ncbi.nih.gov/blast/db/FASTA/env\_nr.gz}}. This database comprises 1384686404 amino acid residues in 6962291 sequences,  11944 being the maximum length. The queries have been extracted from the Swiss-Prot database~\footnote{The Swiss-Prot database is available online at \url{http://web.expasy.org/docs/swiss-prot_guideline.html}} (accession numbers: P02232, P05013, P14942, P07327, P01008, P03435, P42357, P21177, Q38941, P27895, P07756, P04775, P19096, P28167, P0C6B8, P20930, P08519, Q7TMA5, P33450, and Q9UKN1), ranging in length from 144 to 5478. The scoring matrix selected was BLOSUM62, and gap insertion and extension penalties were set to 10 and 2, respectively. 

\subsection{Performance Results}
\label{sec:perf-knl}

% poner en alguna parte porqué no incluimos QP en las SSE y AVX2

\begin{figure}[tb]
\begin{centering}
\includegraphics[width=0.8\columnwidth]{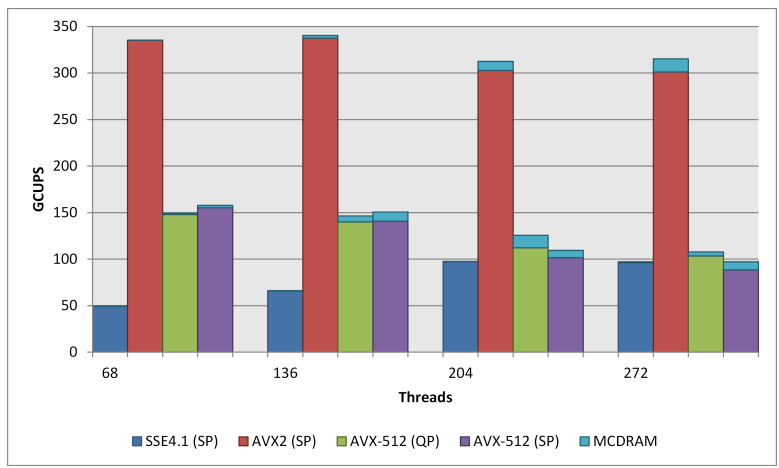}
\par\end{centering}
\caption{\label{fig:threads-instructions}Performance for the different instruction sets used varying the number of threads.}
\end{figure}

\emph{Cell updates per second} (CUPS) is a commonly used performance measure in the Smith-Waterman context, because it allows removal of the dependency on the query sequences and the databases utilized for the different tests. A CUPS represents the time for a complete computation of one cell in matrix \emph{H}, including all memory operations and the corresponding computation of the values in the \emph{E} and \emph{F} arrays. Given a query sequence $Q$ and a database $D$, the GCUPS (billion cell updates per second) value is calculated by:
\begin{equation}
 \frac{|Q| \times |D|}{t \times 10^{9}}\label{eq:gcups}
\end{equation}
where $|Q|$ is the total number of residues in the query sequence, $|D|$ is the total number of residues in the database and \emph{t} is the runtime in seconds~\cite{Rognes2011}.

Figure~\ref{fig:threads-instructions} shows the performance for the different instruction sets used varying the number of threads~\footnote{SSE4.1 and AVX2 versions using \emph{QP} technique were excluded from the analysis to improve figure readability since we found that \emph{SP} scheme always achieved the best performance, as in previous work~\cite{SW-CPE15}}. The best performances are achieved by AVX2 extensions (340.3 GCUPS) followed by AVX-512 (157.8 GCUPS) and, last, SSE4.1 (97.6 GCUPS). As mentioned before, data level exploitation is critical to achieve maximum performance in this application. Even though AVX-512 doubles vectorial width of AVX2 instructions, the lack of low-range integer operations imposes a strong limit to its performance taking into account that almost all alignment scores can be represented using 8-bit integer data. Despite the fact that the SSE4.1 version computes 16 alignments in parallel as the AVX-512 counterparts, the performance of the former is slower compared to the latter. As only one of the VPUs of each core has support for a subset of byte and word SSE instructions, codes that use these operations suffer performance losses.

In relation to the number of threads, AVX2 implementation reaches top performance using 136 threads although performance with 68 threads is very close (just 1\% slower). Similar behaviors are presented with AVX-512 and SSE4.1 intrinsics. In the AVX-512 case, performance with 68 threads is 3\% higher than the corresponding to 136 threads; while SSE4.1 version is slightly better (1\%) employing 204 threads compared to 272 threads.

Lastly, this figure also allows us to evaluate the performance gains obtained by HBM usage. As the entire application fits in the  MCDRAM, we can benefit from placing all data in that memory using the \emph{numactl} utility (without source code modification). In particular, MCDRAM exploitation achieves an average speedup of 1.04$\times$ and a maximum speedup of 1.1$\times$.

\begin{figure}[tb]
\begin{centering}
\includegraphics[width=0.8\columnwidth]{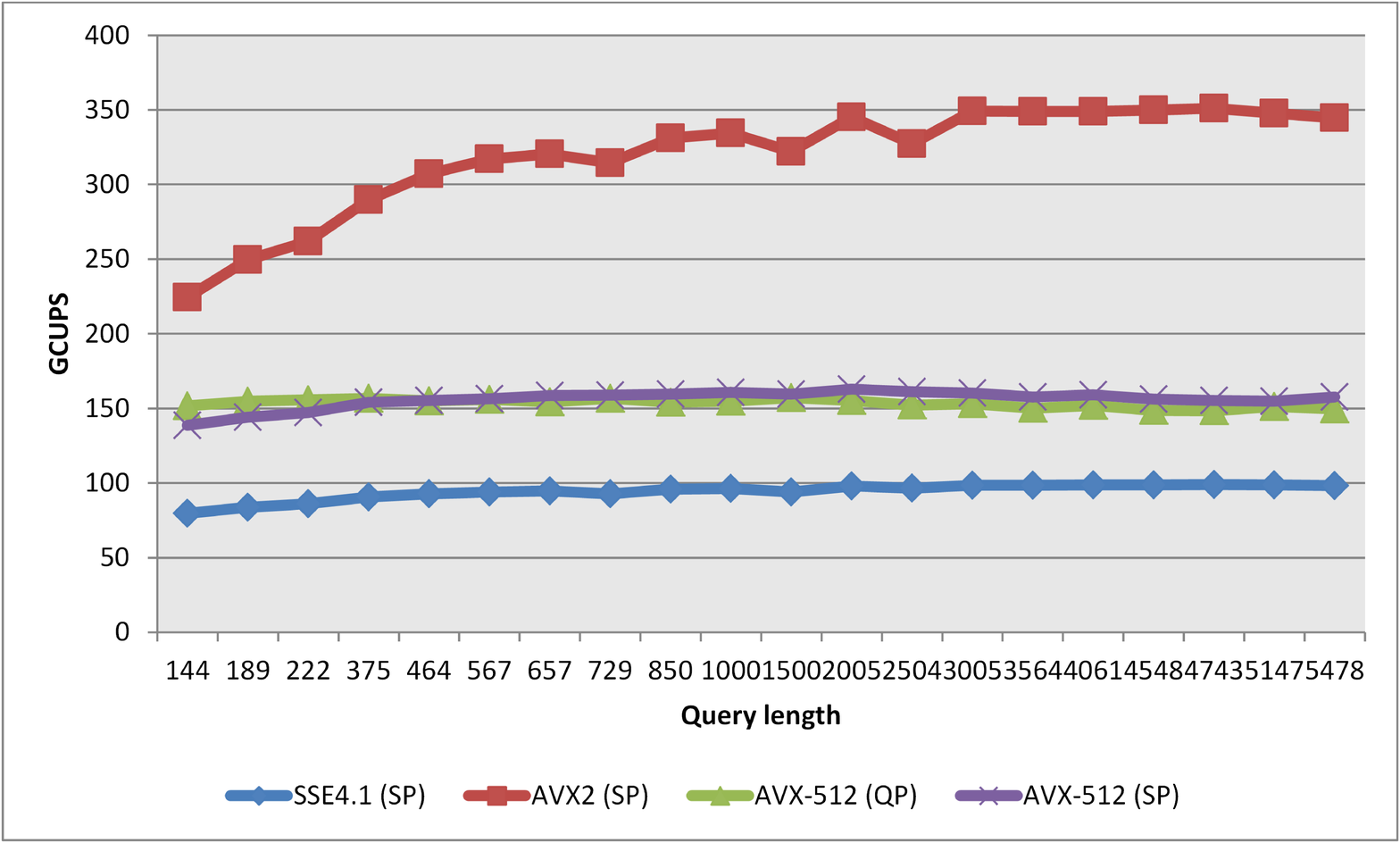}
\par\end{centering}
\caption{\label{fig:workload}Performance evolution varying query length.}
\end{figure}

Figure~\ref{fig:workload} illustrates performance evolution varying query length with the most favorable configuration for each implementation: 204, 136 and 68 threads for SSE4.1, AVX2 and AVX-512 intrinsics, respectively. Also, data is placed in MCDRAM memory. SSE4.1 and AVX-512 implementations have a almost constant performance achievement. As expected, this behavior is motivated by the exploitation of \emph{inter-task} parallelism scheme. AVX2 version achieves an increasing performance tendency that becomes soft with larger query sequences (\emph{m} $\geq$ 2504). For AVX-512, the behavior of \emph{QP} and \emph{SP} differ, observing better performance for short sequences in \emph{QP}. This aspect, also observed in previous research for the Xeon Phi KNC~\cite{SWAPHI,SW-CPE15}, is due to the additional overhead incurred by the \emph{SP} construction, which does not compensate for the indexation benefits in shorter queries. As summary, peak performances achieved are 351.2, 162.8, 157.2 and 98.9 GCUPS for AVX2, AVX-512 (\emph{SP}), AVX-512 (\emph{QP}) and SSE4.1 implementations.

\subsection{Performance Comparison to Parasail Library}
\label{sec:perf-comparison}

\begin{figure}[tb]
\begin{centering}
\includegraphics[width=0.8\columnwidth]{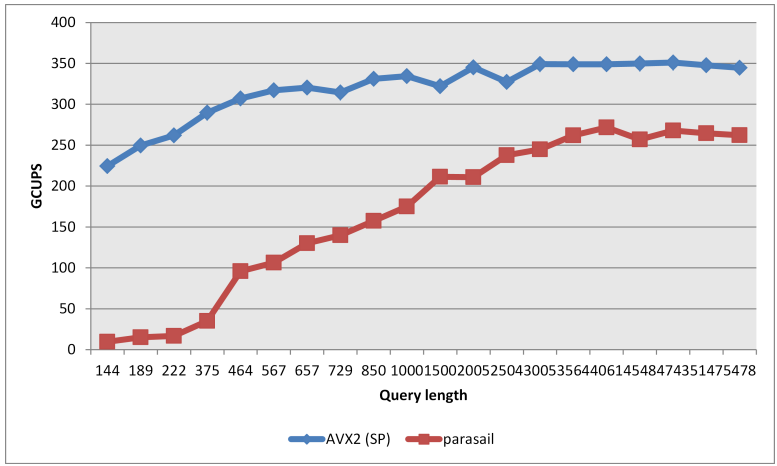}
\par\end{centering}
\caption{\label{fig:comparison}Performance comparison to Parasail library.}
\end{figure}

Finally, we have compared our implementation with the \emph{parasail\_aligner} application included in the Parasail library. As Parasail offers many different alignment scenarios, we tested all and select which reports the best performance rates: \emph{parasail\_sw\_striped\_profile\_avx2\_256\_sat}. This variant is based on the stripped approach for \emph{intra-task} parallelism with AVX2 intrinsics. Besides, it performes \emph{QP} optimization using also 8-bit integer data with overflow checking.

Figure~\ref{fig:comparison} shows the performance comparison between Parasail and our SW version on KNL. Both implementations run with 136 threads and make use of MCDRAM. Parasail \emph{intra-task} approach limits parallel scalability for small aligments. Moreover, our developed version which is based on the \emph{inter-task} and \emph{SP} scheme outperforms Parasail for all query lengths considered. In particular, it runs on average 4.6$\times$ faster highlighting the larger differences for shorter queries.

%% file: conclusions.tex
\section{Conclusions}
\label{sec:conclusion}

The SW algorithm is a critical application in bioinformatics scenario and has become the base of more sophisticated alignment technologies, so its study and acceleration in different platforms has motivated a great interest for the scientific community. In this paper, we have explored SW acceleration on the last generation of Intel's Xeon Phi processors with the KNL architecture. To the best of the authors knowledge, this is the first study of this kind.

Among main contributions of this research we can summarize:
\begin{itemize}
\item Exploitation of low-range integer vectors is crucial to achieve top performance. Even though AVX-512 doubles vectorial width of AVX2 instructions, the latter reach the maximal performance. The lack of this class of AVX-512 instructions in Xeon Phi KNL processors imposes a strong limit to its performance taking into account that almost all alignment scores can be represented using 8-bit integer data.
\item Multi-threading must be carefully evaluated. Different number of threads produced the best results for each instruction set.
\item MCDRAM usage demonstrated to be an effective way to increase performance with practically null programmer intervention. In particular, it produced an average speedup of 1.04$\times$ and a maximum speedup of 1.1$\times$.
\item Peak performances are 351.2, 162.8, 157.2 and 98.9 GCUPS for AVX2, AVX-512 (\emph{SP}), AVX-512 (\emph{QP}) and SSE4.1 implementations.
\end{itemize}

In view of the obtained results, as future works we will consider:

\begin{itemize}
\item Xeon Phi KNL processors offer different cluster and memory modes. We are interested in exploring the \emph{Flat mode} with larger genomic databases that do not fit in MCDRAM, like UniProtKB/TrEMBL. Also, we will evaluate programming and optimization techniques in other available modes as a way to extract more performance.
\item As Xeon Phi KNL processors reported competitive performance, we plan to perform a comparison with other accelerators not only from performance perspective but also from power efficiency point of view.
\item Future Xeon KNL processors will include AVX-512BW set. As this characteristic enables more SIMD parallelism, we see a promising opportunity in accelerating SW database searches on these devices.
\end{itemize}